# Generalized Laws of Refraction and Reflection at Interfaces between Different Photonic Artificial Gauge Fields

Moshe-Ishay Cohen[*1,2], Christina Jörg[*3,4], Yaakov Lumer[1,2], Yonatan Plotnik[1,2], Erik H. Waller[3], Julian Schulz[3], Georg von Freymann[3,5] and Mordechai Segev[1,2]

[1]Physics Department, Technion – Israel Institute of Technology, Haifa 32000, Israel
[2]Solid State Institute, Technion – Israel Institute of Technology, Haifa 32000, Israel
[3]Physics Department and Research Center OPTIMAS, TU Kaiserslautern, 67663 Kaiserslautern, Germany
[4]Department of Physics, The Pennsylvania State University, Pennsylvania 16802, USA
[5]Fraunhofer Institute for Industrial Mathematics ITWM, 67663 Kaiserslautern, Germany
* Equal contribution

msegev@technion.ac.il

Abstract

**Artificial gauge fields enable extending the control over dynamics of uncharged particles, by engineering the potential landscape such that the particles behave as if effective external fields are acting on them. Recent years have witnessed a growing interest in artificial gauge fields that are generated either by geometry or by time-dependent modulation, as they have been the enablers for topological phenomena and synthetic dimensions in many physical settings, e.g., photonics, cold atoms and acoustic waves. Here, we formulate and experimentally demonstrate the generalized laws of refraction and reflection from an interface between two regions with different artificial gauge fields. We use the symmetries in the system to obtain the generalized Snell law for such a gauge interface, and solve for reflection and transmission. We identify total internal reflection (TIR) and complete transmission, and demonstrate the concept in experiments. Additionally, we calculate the artificial magnetic flux at the interface of two regions with different artificial gauge, and present a method to concatenate several gauge interfaces. As an example, we propose a scheme to make a gauge imaging system – a device that is able to reconstruct (image) the shape of an arbitrary wavepacket launched at a certain position to a predesigned location.**

# Introduction

Snell's law and the Fresnel coefficients are the cornerstones of describing the evolution of electromagnetic waves at an interface between two different media. By cascading several such systems, each with its own optical properties, it is possible to design complex structures which give rise to various important devices and systems such as lenses, waveguides[1], resonators, photonic crystals[2] and even localization phenomena – when random interfaces are involved[3]. The behavior of waves in the presence of an interface can exhibit quite interesting features, e.g.: total internal reflection (TIR), back-refraction for negative-positive refraction index interfaces[4,5]; and even states that are confined to the interface itself like Tamm and Shockley states[6,7], plasmon polaritons[8,9], Dyakonov states[10,11] and topological edges states[12–14]. Traditionally, the Fresnel equations describe reflection and transmission of electromagnetic waves at an interface separating two media with different optical properties. These can be two materials with different permittivities or two different periodic systems (photonic crystals) made up from the same material, e.g., an interface between two dissimilar waveguide arrays[15]. However, an interface can also separate between two optical systems that differ only by the artificial gauge fields created in them. Generally, such a "gauge interface" alters the dispersion curve at either side of the interface, hence it must affect the transmission and reflection by the interface.

Gauge fields (GF) are a basic concept in physics describing forces applied on charged particles. Artificial GF are a technique for engineering the potential landscape such that neutral particles will mimic the dynamics of charged particles driven by external fields. With the advent of particle-wave duality, artificial GF have been demonstrated to act on photons[16–19], cold atoms[20,21], acoustic waves, etc. These artificial GF are generated either by geometry[17] or by time-dependent modulation[18] of system parameters. With the growing interest in topological systems[22], which necessitate GF[23,24], it was suggested that the interface between two regions of the same medium, but with different GF in each region, can create an effective edge. In these systems both sides of the interface have the same basic dispersion properties, altered only by applying a different GF at each side. Such a gauge edge was employed to demonstrate analogies to the Rashba effect[25], optical waveguiding[26,27], topological edge states[28,29] and back-refraction[30]. In the presence of a different GF on either side of the interface, the trajectories of waves crossing from one side to the other is governed by the symmetries in the system, which are expected to result in an effective Snell's law, whereas the reflection and transmission coefficients arise from the specific boundary.

Here, we demonstrate theoretically and experimentally the effective Snell law governing the reflection and transmission of waves from an interface between regions of the same photonic medium differing only by the artificial gauge fields introduced on either side. We show how the reflected and transmitted waves change their transverse momenta according to the interfacial change in the gauge field, and demonstrate the total internal reflection and complete transmission. Subsequently, we provide an approximate calculation for the Fresnel coefficients for our example, and explain how to generalize the concepts. Finally, we show how to concatenate several "gauge interfaces", and propose a design for a gauge-based imaging system – a device made up of several interfaces between different gauge fields, acting to reconstruct (image) an arbitrary paraxial input wavepacket at a predetermined plane.

**Results**

For simplicity, consider first a simple system constituting an artificial gauge interface: two 2D arrays of evanescently coupled waveguides, where the waveguides in each array follow a different trajectory along the propagation axis $z$ (Fig. 1). This model system serves to explain the ideas involved, which are later on generalized. The GF in our system is a direct result of the trajectories of the waveguides, and does not require any temporal modulation of the materials at hand. The upper array experiences a constant tilt in the $x$-direction along the propagation axis $z$, with a paraxial angle $\eta$, such that $x(z) = x' + \eta z$ (paraxiality allows $\sin \eta \simeq \eta$, and $x'$ is the original $x$-position), while the lower is tilted by $-\eta$ (Fig 1(a)).

These two arrays combined exhibit different artificial GFs that cannot be gauged away by a coordinate transformation. The propagation of light in this structure is described by the paraxial wave equation

$$i\partial_z \psi(\vec{r}) = -\frac{1}{2k_0}\nabla_\perp^2 \psi(\vec{r}) - \frac{k_0}{n_0}\Delta n(\vec{r})\psi(\vec{r}). \quad (1)$$

Here, $\psi$ is the envelope of the electric field, $k_0$ is the optical wavenumber inside the bulk material, $n_0$ is the ambient refractive index, $\Delta n(\vec{r}) = n(\vec{r}) - n_0$ gives the relative refractive index profile and $\nabla_\perp^2 = \partial_x^2 + \partial_y^2$. Equation (1) is mathematically equivalent to the Schroedinger equation, where $z$ plays the role of time, and $\Delta n(\vec{r})$ plays the role of the potential. This analogy between the paraxial wave equation and the Schroedinger equation has been used many times in exploring a plethora of fundamental phenomena, ranging from Anderson localization[31] and Zener tunneling[32] to non-Hermitian potentials[33] and Floquet topological insulators[23].

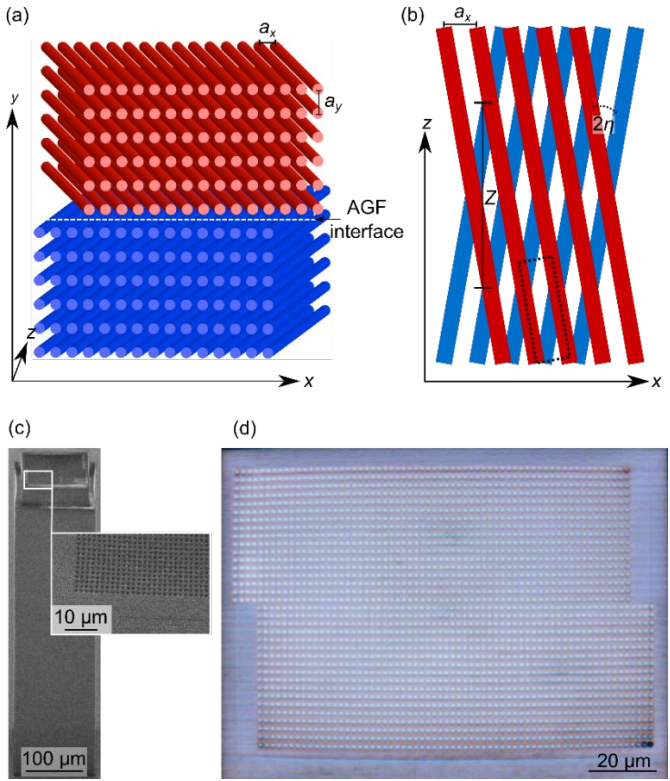

**Figure 1: Sketch of our artificial gauge interface.** Two rectangular arrays of waveguides (red: upper array, blue: lower array) are stacked on top of each other, creating an artificial GF interface in the $y$-direction. The waveguide arrays are tilted by $2\eta$ with respect to each other. Otherwise, the parameters for the two arrays are identical. (a) Front view. (b) Top view. The dashed box represents one unit cell in the $x$-$z$-direction in the upper array. (c) SEM image of the inverse fabricated waveguide sample from the side. The inset shows a magnified region to visualize the hollow waveguides. (d) Microscope image of the infiltrated sample from the top.

The basic building block in our system is a two-dimensional array of evanescently coupled straight waveguides, i.e., $\Delta n(\vec{r})$ is a periodic function in both $x$ and $y$, with periods $a_x$ and $a_y$, respectively, such that each unit cell consists of a single waveguide.

Consider first an array where the trajectories of all the waveguides are in the $z$-direction. Following coupled mode theory[25], the spectrum of light propagating in such a 2D array of waveguides is given by

$$\beta(k_x, k_y) = \beta_0 + 2c_x \cos(k_x a_x) + 2c_y \cos(k_y a_y) \quad (2)$$

where $\beta$ is the propagation constant of an eigenmode, defined by $\psi(x, y, z) = \psi_0(x, y)e^{-i\beta z}$, $k_x$ and $k_y$ are the spatial momenta of the mode in the $x$- and $y$-directions, $c_x$ and $c_y$ are the coupling strengths between adjacent waveguides in the $x$- and $y$-directions (taken to be real negative numbers according to standard solid states notation) and $\beta_0$ is the propagation constant of the guided mode in a single isolated waveguide. Consider now an array of waveguides tilted at an angle $\eta$ with respect to the $z$-axis, such that the waveguides follow a trajectory defined by $x - \eta z = $ constant. The dynamics in an array of tilted waveguides is expressed by an artificial GF, given by the effective vector potential $\vec{A}(z) = -k_0 \eta \hat{x}$, with the following spectrum[25,27]:

$$\beta_\eta(k_x, k_y) = \beta_0 + 2c_x \cos((k_x - k_0\eta)a_x) + \eta k_x - \frac{1}{2}k_0\eta^2 + 2c_y \cos(k_y a_y) \quad (3)$$

The shift of $k_0\eta$ in the cosine is the compensation due to Galilean transformation of the waveguides. The linear $\eta k_x$ shift term appears because the spectrum in Eq. (3) is expressed in the laboratory frame, and not in the frame of reference in which the waveguides are stationary. The constant offset $\frac{1}{2}k_0\eta^2$ results from the effective elongation of the optical path inside the tilted waveguides.

Such a linear tilt of a waveguide array is, in itself, a trivial gauge field, as we can annihilate its effects by changing the frame of reference to the co-moving frame, i.e., a linear coordinate transformation of the entire system can gauge it out. This can also be understood by looking at the arising effective magnetic field $\vec{B} = \vec{\nabla} \times \vec{A}$, which turns out to be zero for a constant vector potential $\vec{A}$. In order to have a non-trivial gauge, we need the effective gauge field to be non-uniform (i.e., have either space or time dependence). Such a non-trivial gauge is achieved by coupling two 2D arrays, each with a different tilting angle, and therefore a different gauge. Then, it becomes impossible to gauge away the effect of the tilt, when we combine two of such fields with different tilts. There is no coordinate system in which both arrays would simultaneously be untilted.

With this in mind, consider a two-dimensional array of evanescently coupled waveguides, divided into two regions – top and bottom as shown in Fig. 1(a). The rows of waveguides at the top and bottom regions are identical in every parameter except for the tilt. The overall GF is then given by

$$\vec{A}(\vec{r}) = (2\Theta(y) - 1)\eta k_0 \hat{x} \tag{4}$$

where $\Theta(y)$ is the Heaviside function, which is 1 when $y > 0$ and zero otherwise. This vector potential gives an effective magnetic field $\vec{B} = -2\delta(y)\eta k_0 \hat{z}$ with a $\Phi_B = -2\eta k_0 a_x$ magnetic flux through a unit cell at the interface.

The different gauge in each subsystem results in a different band-structure (dispersion relation) for each subsystem. The system has discrete symmetries in both the $x$- and the $z$-directions, with a period of $a_x$ and $\frac{a_x}{2\eta}$, respectively (see Supplementary Information). Each of these dictates a conservation law for its respective momentum (up to $2\pi$ over the period), leaving only $k_y$ to be modified as the wave crosses between the two regions. Thus, launching an eigenmode (a Bloch wave) with a defined wavevector $(k_x, k_{y,\text{inc}})$ in one side of the system will result in refraction and reflection of the wave, upon incidence at the interface. According to the $x$- and $z$- translational symmetries of the joint lattice, the wavenumber in the second half plane will have to satisfy:

$$\beta_\eta(k_x, k_{y,\text{inc}}) = \beta_{-\eta}(k_x, k_{y,\text{tran}}) \tag{5}$$

where $k_{y,\text{inc}}$ is the $y$-wavenumber of the incident beam, $k_{y,\text{tran}}$ is the $y$-wavenumber of the transmitted beam, $\beta_\eta(k_x, k_{y,\text{inc}})$ is given by Eq. (3). Equation (5) acts as a generalized Snell law for an interface between two regions of the same medium but with different artificial gauge fields at each side, in the specific realization of titled photonic lattices.

Note that Eq. (5) is general and valid for any interface that satisfies the symmetries in $x$ and $z$ (the plane normal to the interface), even for a uniform dielectric medium. The main difference between refraction from a dielectric interface and refraction from an AGF interface is in the dispersion relation, that is, the relation between the propagation constant $\beta$ and the frequency. In uniform dielectrics, a plane wave (which is an eigenmode of the medium) obeys $\beta_{\text{dielectric}} = \sqrt{\left(\frac{n\omega}{c}\right)^2 - k_x^2 - k_y^2}$. On the other hand, for an AGF medium, the dispersion can be designed to almost any desired relation[34,35]. In the specific case of tilted waveguide arrays, the dispersion relation is given by Eq. (3). The ability to design the dispersion allows for interesting dynamics such as negative group velocity for some $k_x$ values; for example, $2c_x a_x \sin((k_x - k_0\eta)a_x) < -\eta$ for the case of tilted waveguide arrays. Engineering the dispersion relation also makes it possible to cancel diffraction in one of the directions, as we suggest in the discussion section (Figure 7). Notably, uniform dielectric interfaces require materials with different optical properties at each side of the interface, whereas an AGF interface can be engineered even when both sides have the same optical properties (up to their gauge), achieving refraction using the same materials and of the same composition and configuration (periodicity).

Figure 2 shows the band structures for the upper (red) and lower (blue) waveguide array. Depicted in three dimensions (Fig. 2(b)), we notice the sinusoidal shape of the dispersion along $k_x$ as well as along $k_y$. The projection onto the $k_x$-component (Fig. 2(a)), however, helps us display the $k_y$-conversion between the two arrays. In the projected band structure, each band represents the values of $\beta$ (which plays the role of energy in the analogy to the Schrödinger equation) for all values of $k_y$ associated with that band (see Supplementary Information for discussion on band replicas arising from the periodicity of the structure in $x$ and in $z$). As an example, the solid lines in Fig. 2(a) indicate the values of $\beta$ associated with some specific $k_y$. Notice that the $k_x$-range for total reflection (dashed vertical lines in Fig 2(a)) is not necessarily symmetric around $k_x = 0$. From this figure, it may seem that $\beta$ is not periodic in $k_x$, but a closer look on the symmetries in the system reveals that the periodicity is maintained (see discussion in the Supplementary Information). Figure 2(c) displays a contour plot of equi-$\beta$ as a function of $k_x$ and $k_y$. The group velocity of a wavepacket at each point is perpendicular to an equi-$\beta$ contour that goes through that point.

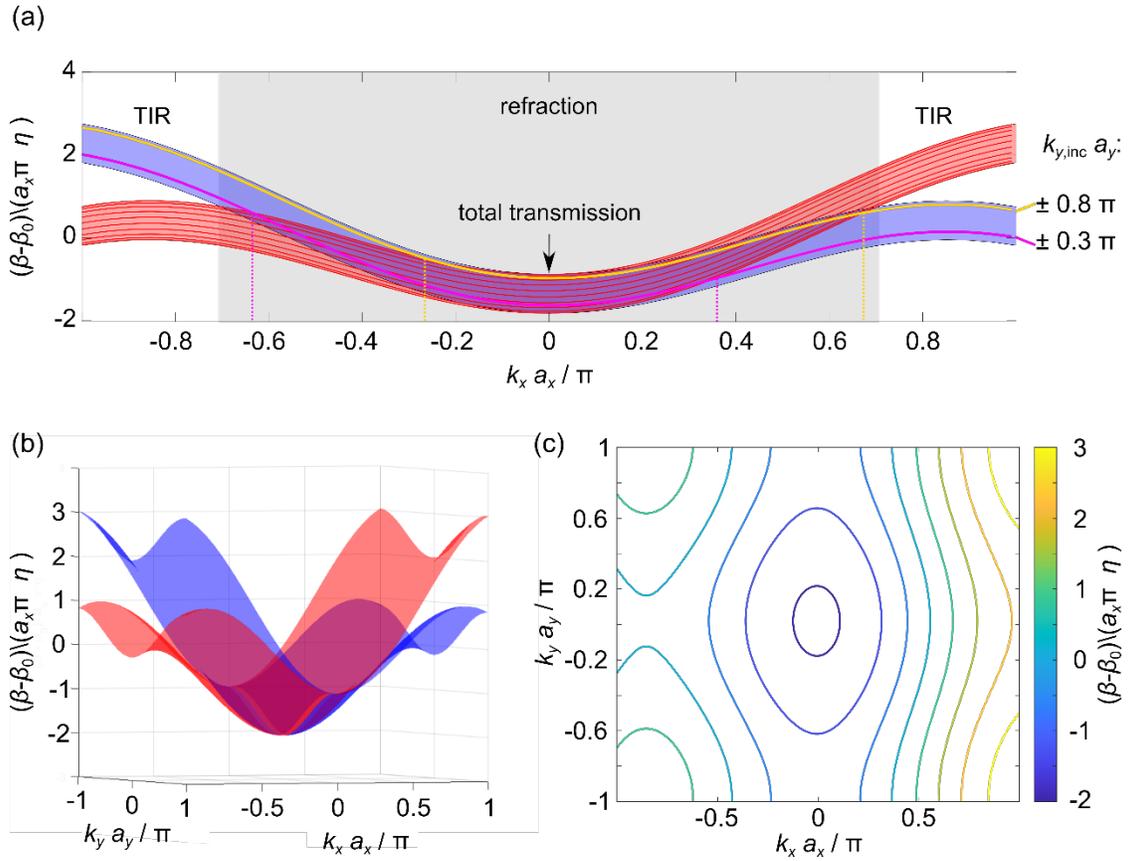

**Figure 2: Dispersion relations for the upper (red) and lower (blue) arrays.** (a) Projection of $\beta$ as a function of $k_x$. The solid lines mark some specific values of $k_y a_y$. In the $k_x$-range where the two bands overlap, any wavepacket traveling across the artificial gauge interface undergoes refraction according to the generalized Snell law (shaded grey region). Outside this range, total internal reflection (TIR) occurs. Notice that the $k_x$-range for total reflection (unshaded region) is not necessarily symmetric around $k_x = 0$ (see vertical dashed lines). The parameters here are as in the experiment: $r = 0.52$ μm, $a_x = 1.69$ μm, $a_y = 2.15$ μm, $\eta = 0.0093$, $\lambda = 700$ nm. (b) Full 3D dispersion relation $\beta$ as a function of $k_x$ and $k_y$. (c) Contour plot of the equi-$\beta$ over $k_x$ and $k_y$ for the red band. Each line represents an equi-$\beta$ contour. The group velocity of a wavepacket at each point is perpendicular to an equi-$\beta$ contour that goes through that point.

When a wavepacket crosses the artificial GF interface between the lower array and the upper array, $\beta$ is conserved, such that $\beta_\eta(k_x, k_{y,\text{inc}}) = \beta_{-\eta}(k_x, k_{y,\text{tran}})$, as well as the transverse wavenumber $k_{x,\text{inc}} = k_{x,\text{tran}} = k_x$. Graphically, this means that, at this value of $\beta$, the red and the blue bands in Fig. 2(a) overlap. Then, the quasi-energy $\beta$ of the red band may belong to a different $k_y$ than that of the blue band, thus $k_{y,\text{inc}} \neq k_{y,\text{tran}}$ (see solid colored lines in Fig. 2(a)). Therefore, when the light crosses the artificial GF interface, $k_{y,\text{inc}}$ has to change according to:

$$\cos(k_{y,\text{tran}} a_y) - \cos(k_{y,\text{inc}} a_y) = \frac{c_x}{c_y}\left[\cos((k_x + k_0\eta)a_x) - \cos((k_x - k_0\eta)a_x)\right] - \eta\frac{k_x}{c_y} \quad (6)$$

We identify three different regimes, which depend on $k_x$:

1) Total internal reflection (TIR): when $k_x$ is such that the blue and red bands do not intersect, hence no coupling from the upper to the lower array (and vice versa) is possible. Consequently, the wavepacket is reflected completely (see Fig. 3(i)).

2) Perfect transmission for $k_x = 0$: A wavepacket crosses the interface between the two arrays without changing its wave vector components while allowing all the light to be transmitted through the interface(see Fig. 3(a,c)).

3) Refraction and reflection for all other values of $k_x$: regions where the red and blue bands intersect, but for different $k_{y,\text{inc}}$ and $k_{y,\text{tran}}$. As the "energy" $\beta$ and the transverse wavenumber $k_x$ are conserved, $k_y$ has to change upon crossing the interface, resulting in both a refracted wave and a reflected wave (Fig. 3(d,f)).

Examples for the dynamics of waves in these three regimes are given in Fig. 3, which shows the results of direct simulations of Eq. (1) (using the OptiBPM code), with parameters corresponding to those used in the experiments. The figure shows the intensity of optical beams at three different propagation planes, $z$. We probe the three different regimes by launching input beams with a set $k_{y,\text{inc}} a_y = 0.5\,\pi$, and selecting $k_x a_x$ corresponding to total transmission ($k_x a_x = 0$), refraction and reflection ($k_x a_x = 0.4\,\pi$), and total internal reflection ($k_x a_x = 0.8\,\pi$). Upon excitation at $z = 0$ μm (Fig. 3(a,d,g)), the beams travel towards the interface (indicated by the dashed white line), reaching it approximately after $z = 600$ μm (Fig. 3(b,e,h)). For the input beam with $k_x a_x = 0$, the beam is completely transmitted across the interface (Fig. 3(c)) without any reflection. After passing through the interface, the beam disperses (diffracts) strongly in the $x$-direction. For the input beam with $k_x a_x = 0.4\,\pi$, part of the beam is reflected by the interface, returning to the lower array, while part of it is refracted, as indicated by the change in the slope of the arrow (Fig. 3(f)). For the beam with $k_x a_x = 0.8\,\pi$, the beam undergoes total reflection, never crossing the interface (Fig. 3(i)).

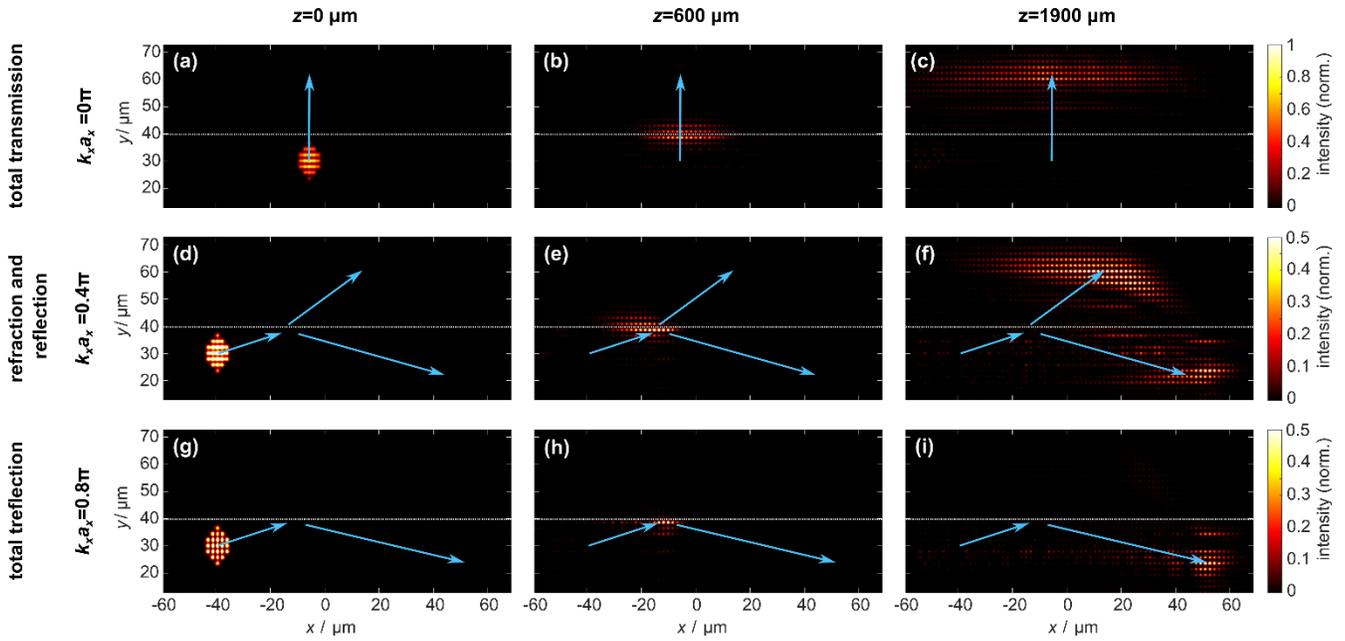

**Figure 3: Simulated dynamics of beams in three regimes:** Total transmission (upper row), refraction and reflection (middle row) and total internal reflection (bottom row). Each panel shows the numerically calculated intensity distributions at three different z-values along the propagation: input facet (left column), $z = 600$ μm (middle column) and $z = 1900$ μm (right column). The dashed white line indicates the location of the gauge interface. The initial y-wavenumber is always $k_{y,\text{inc}} a_y = 0.5\ \pi$. (a-c) An input beam with $k_x a_x = 0$ is transmitted completely across the interface. (d-f) An input beam with $k_x a_x = 0.4\ \pi$ is partially reflected and partially refracted at the interface. The refraction is highlighted by the change in the trajectory. (g-i) An input beam with $k_x a_x = 0.8\ \pi$ undergoes total reflection, never crossing the interface. The intensity is normalized separately in each panel, while the intensity in the second and third row is enhanced for better visibility. The arrows are guides to the eye and indicate the approximate trajectories of the respective beams.

Having used symmetry and the dispersion relation to find the general laws of refraction and reflection at a gauge interface, the next step is natural: finding the coefficients for reflection and transmission. However, just like the Fresnel coefficients at a dielectric interface, this calculation is system specific, and depends in detail on the interface between the two regions. That is, unlike the Snell-like law, the Fresnel coefficients cannot be generalized (conservation of power yields a relation between the absolute values of the Fresnel coefficients, but to obtain the coefficients one must also employ continuity at the interface). With this in mind, we calculate the Fresnel-like coefficients for our example of a gauge interface constructed from titled waveguide. We use an approximate model[27] for the coupling between the two sections, and derive an approximate formula for the coefficients. The details of the calculation are presented in the Supplementary Information. Figure 4 shows the Fresnel-like coefficients for our gauge interface, for parameters corresponding to those shown in Fig. 3. We plot the amplitude and phase of the transmitted and reflected parts for $k_{y,\text{inc}} a_y = 0.4\pi$ as a function of $k_x$, and for $k_x a_x = 0.4\pi$ as a function of $k_{y,\text{inc}}$ (Figs. 4(c) and (d) respectively). As explained above, the Fresnel coefficients are highly dependent on the specifics of the interface, hence a different model for the gauge field interface will yield different coefficients. However, the Snell-like law of refraction will not change, as it depends only on symmetry and the dispersion relations at both sides of the interface.

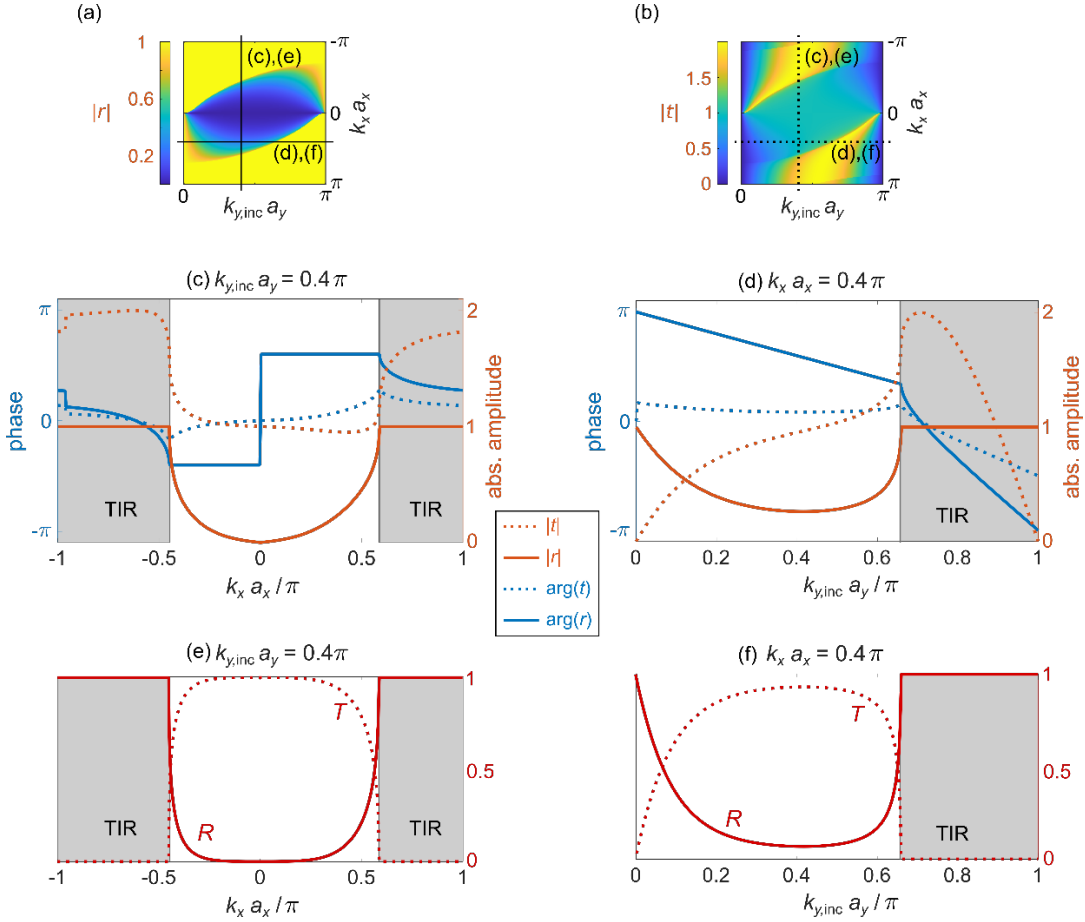

**Figure 4: Fresnel coefficients**: Reflection amplitude (a) and transmission amplitude (b) across the entire $k_x$ and $k_y$ range. The line cuts on the right hand side show the phase (blue line) and amplitude (red line) of the reflection coefficient $r$ (solid line) and the transmission coefficient $t$ (dashed line) along the cuts indicated in (a) and (b): for $k_{y,\text{inc}} a_y = 0.4\pi$ (c) and $k_x a_x = 0.4\pi$ (d). Note that $|t|$ can be bigger than 1 for some $(k_x, k_{y,\text{inc}})$. (e) and (f) show the reflectance $R = |r|^2$ (solid line) and transmittance $T = \text{Re}\left\{\frac{v_{gy,\text{tran}}}{v_{gy,\text{inc}}}\right\}|t|^2$ (dashed line) for the same wave vectors as in (c) and (d). One can easily convince oneself that the total energy flux is indeed conserved, as $R + T = 1$ holds for all regions.

To demonstrate the generalized Snell's law in experiments, we fabricate sets of tilted optical waveguides corresponding to the system described in Fig. 1. The samples are fabricated using direct laser writing to create hollow waveguides which are subsequently infiltrated with higher index material (Fig. 1(c) and (d)). For details on the fabrication see Ref.[36]. The experimental set up is sketched in Fig 5. We reflect a 700 nm laser beam off a spatial light modulator (SLM) to excite a Bloch mode with a given $k_{y,\text{inc}}$ while exciting the entire first Brillouin zone in $x$ (i.e., $-\pi \leq k_x a_x \leq \pi$). To do that, the beam reflected off the SLM[37] is shaped such that after a Fourier transform (by a microscope objective) it consists of 5 lobes, with their phase forming a linear ladder, commensurate with the chosen Bloch mode, while in $x$ the beam is simply focused into a single row of waveguides. The beam passes through the sample, and the output facet is imaged onto a camera. Since we are interested in investigating the passage through the gauge interface, we scan the value of $k_{y,\text{inc}}$ by changing the relative phase between the lobes, while exciting the entire first Brillouin zone in $x$. Figure 5 shows a false color photograph of the input beam on top of the sample, with the interface marked by the dashed line. The light propagates across the interface, and we measure the intensity of the refracted and reflected waves at the output facet of the sample, as well as the intensity at the Fourier plane (obtained at the focal plane of a lens), which corresponds to the spatial power spectrum.

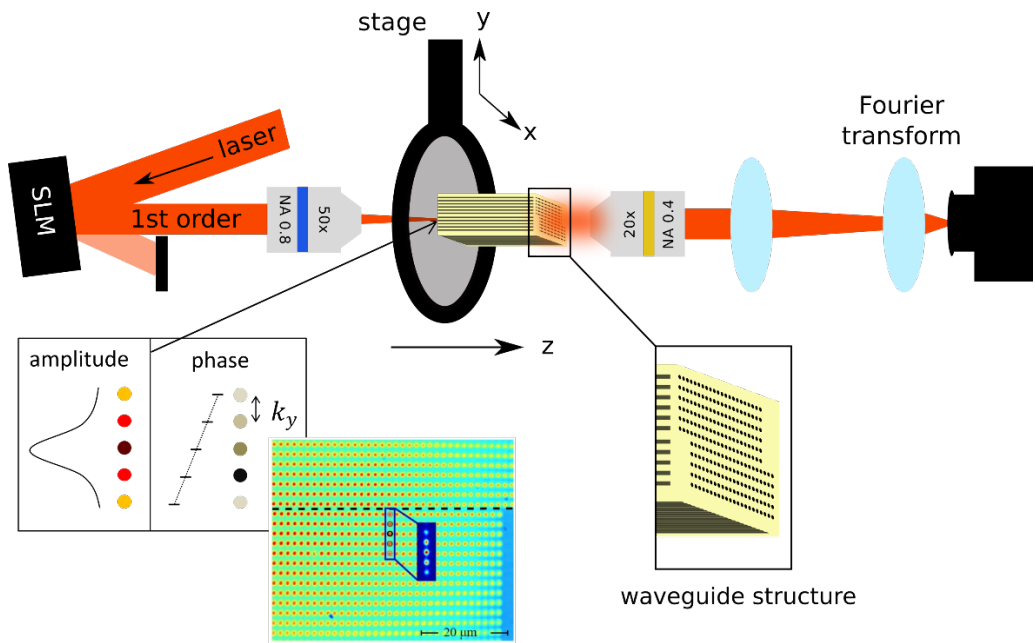

**Figure 5: Experimental setup**: a laser beam (wavelength 700 nm) is reflected off an SLM, which imprints a specific phase and amplitude pattern onto the beam. To shape the amplitude, while using a phase-only SLM, we overlay the phase pattern on the SLM with a blazed grating that shifts parts of the reflected light into the first diffraction order[37]. After Fourier transform by an objective, the beam consists of five spots with a phase difference of $k_{y,inc} a_y$ between each of them and a Gaussian amplitude envelope (inset). These spots are focused onto a row of waveguides below the interface, as shown by the false color photograph of the beam on top of the sample, with the interface marked by the dashed line. The light then propagates along $z$, interacts with the interface, and exits the sample after a propagation of $z = 725$ μm. The intensity distribution at the output facet is imaged onto a camera, as well as its spatial power spectrum (obtained by inserting an extra lens in focal distance to the camera).

Figure 6 shows the spatial power spectrum (Fourier space intensity) of the waves exiting the sample, for an input wave at different $k_{y,inc}$ values, but always launched at the same position in $y$, along with a comparison to beam-propagation simulations. The analytically calculated values for $k_{y,tran}$ obtained by Eq. (6) are marked by the green and blue dots on top of the experimental and simulated results. The beam travels towards the artificial GF interface with a group velocity $v_{gy}$, obtained by taking the derivative with respect to $k_y a_y$ from the dispersion relation in Eq. (3). In the first row (Fig. 6(a),(b)) the beam is launched to move away from the interface ($k_y a_y = -0.6\,\pi$) such that the beam never reaches the interface, hence the output beam has the same spatial spectrum as the input beam. As Fig. 6(a,b) show, the power spectrum of the output beam is located around the same $k_y a_y = -0.6\,\pi$ as the input beam. In the second row (Fig. 6(c),(d)), the beam is launched with $k_y a_y = 0.1\,\pi$. As these panels show, the output beam is split: for $-0.1\,\pi < k_x a_x < 0.6\,\pi$ the beam is transmitted and displays distinct refraction, according to the generalized Snell's law expressed by Eq. (6), while in the regions beyond this range the beam experiences TIR. Notice the prominent asymmetry between the minimal and maximal $k_x$ boundaries between the regions of transmission and TIR. In the experiment (c), the measurement shows the results only partially, as the group velocity in $y$ for $k_y a_y = 0.1\,\pi$ is very low, and the beam has only partially passed the interface even upon reaching the output facet of the sample. In the third row, Fig. 6(e),(f), the beam is launched with $k_y a_y = 0.5\,\pi$. For $-0.5\,\pi < k_x a_x < 0.5\,\pi$ the beam is transmitted, while beyond this range the beam experiences TIR. Here the asymmetry between the minimal and maximal $k_x$ boundaries between transmission and TIR is less significant. The experiment (Fig. 6(e)) captures both the refraction and TIR regions. In the 4$^{th}$ row, Fig. 6(g),(h) the beam is launched with $k_y a_y = \pi$. For $-0.7\,\pi < k_x a_x < -0.1\,\pi$ the beam is transmitted. Beyond that domain the beam experiences TIR. Here, both the incident and the reflected beams have the same $|k_y|$. This case is similar to prism coupling at a grazing angle, in order to couple a beam into a waveguide where it will be bound by TIR.

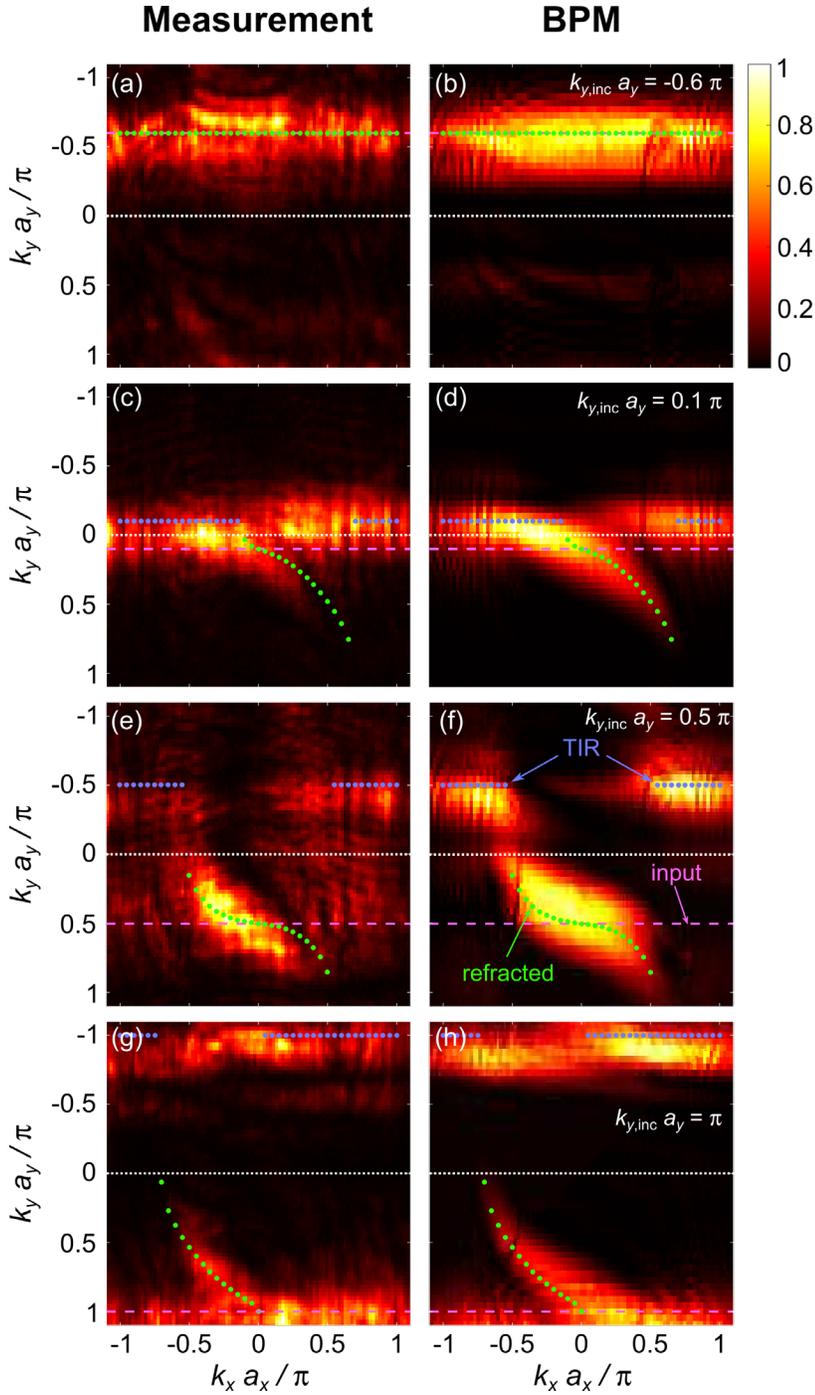

**Figure 6: Refraction and reflection by an artificial gauge interface.** Spatial power spectrum (intensity in the Fourier space), for an input wave at different values of $k_{y,\text{inc}}$ always launched at the same position in $y$. The left and right columns depict the experimental and simulated results, respectively, while the dots show the analytically calculated values from Eq. (6). To obtain each experimental image, we average over 10 measurements with different input positions along $x$. For $k_{y,\text{inc}} a_y = -0.6\,\pi$ the beam travels away from the interface and does not refract at all (a-b), while for the other panels, we see both partial refraction and reflection.

Comparison of the experimental measurements and numerical calculations with the analytical equation (6) (dots in Fig. 6) shows a good agreement for the refracted part. As expected, the $k$-distribution obtained is broader than the analytic curve due to finite size effects. Namely, the input beam with $k_{y,\text{inc}}$ has a finite width (see Fig. 5) of 5 waveguides. The more waveguides are excited in real-space, the smaller the width of the $k$-component in Fourier-space. However, we do not want the input pattern to excite waveguides in the other array across the artificial GF interface, hence we have to limit the size of the input beam. Also, the center position of the input pattern needs to be close enough to the artificial GF interface, such that the beam can travel across the artificial GF interface in the given propagation distance. Therefore, we need to limit the number of excited waveguides. For an excitation of five waveguides the width is $\Delta k_{y,\text{inc}}\, a_y \approx 0.4\, \pi$ (see Fig. 6). As the same number of illumination spots is chosen in the experiments, the numerical calculation reflects the experimental conditions very well. Altogether, the comparison of the measurements, simulations, and the analytic expression shows good agreement with the expected $k_{y,\text{tran}}$-distribution. A movie showing the complete measurements set can be found in Supplementary Movie #1.

**Discussion**

Having demonstrated the Snell law for refraction and reflection at an interface between two different artificial GF, we move on to concatenate several gauge interfaces and construct devices. As an example highlighting the possibilities that refraction by artificial GF allows, we design a gauge-based imaging system. Realizing such a system with arrays of tilted waveguides, we design a scheme that maps any (arbitrary) input wavepacket at the input facet to the output facet. In the scheme based on waveguide arrays, this corresponds to a system with different rows of waveguides tilted at different angles (Fig. 7(a)), mapping an input state from row $y_0$ to row $y_{\text{image}}$. For every row of waveguides with a tilt $\eta(y)$ positioned at $y$, we find the propagation constant $\beta_{\eta(y)} = \beta_0 + 2c_x \cos\left((k_x - k_0\eta(y))a_x\right) + \eta(y)k_x - \frac{1}{2}k_0\eta^2(y)$. To produce an image, we need that the phase accumulation by all components should be identical. Thus, we require that, when a wavepacket diffracts along $y$ and $z$ from the input row $y_0$ to the output row $y_{\text{image}}$, the cumulative phase accumulation for each of its $k_x$ constituents will be the same. Therefore, the value of $\int_0^{y_{\text{image}}} \beta_{\eta(y)}(k_x)\, dy$ should not depend on $k_x$. This can be written as:

$$\frac{d}{d(k_x)} \int_0^{y_{\text{image}}} \beta_{\eta(y)}(k_x)\, dy = 0 \,. \tag{7}$$

This requirement can be fulfilled in a 2D array of straight waveguides, where each row ($y$) is tilted at a different angle, such that $\eta(y) = \eta_0 \sin\left(\frac{2\pi y}{y_{\text{image}}}\right)$. The requirement in Eq. (7) can be expressed by:

$$\int_0^{2\pi} \exp(-ik_0\eta_0 a_x \sin(y'))\, dy' = J_0(k_0\eta_0 a_x) = 0 \tag{8}$$

where $J_0$ is the zeroth-order Bessel function and $y' = \frac{2\pi y}{y_{\text{image}}}$ is now unitless. In other words, by designing the tilt of each row properly, an arbitrary wavepacket $f(x)$ at row $y = 0$ is reproduced at row $y_{\text{image}}$ (apart from a global phase). Figure 7(a) shows a sketch of the gauge-imaging waveguide structure. Each row has a different tilting angle $\eta(y) = \eta_0 \sin\left(\frac{2\pi(y-y_0)}{y_{\text{image}}-y_0}\right)$. Fig. 7(b) is comparing the amplitude and phase of the launched wavepacket at $y = y_0$ and $z = 0$ (pink and light blue) to the imaged one at $y = y_{\text{image}}$ and $z = 1000$ μm (red and dark blue), to reveal that the final and initial wavepackets are essentially the same. One should notice that the beam diffracts in $y$ as it propagates along $z$, hence the imaging is one-dimensional, for the field distribution in $x$ only. Therefore, the intensity reaching the row at $y_{\text{image}}$ is limited by the 1D discrete diffraction in $z$, such that the intensity at each row (neglecting back reflections) is given by $\left|J_{\frac{\Delta y}{a_y}}(f(z))\right|^2$, where $f(z)$ is a function of $z$ that depends on the details of the coupling and $J_{\frac{\Delta y}{a_y}}$ is the Bessel function of the order of the row number $\frac{\Delta y}{a_y}$. In our simulated example, there are 29 rows of waveguides between $y_0$ and $y_{\text{image}}$ so the maximal intensity that can propagate to $y_{\text{image}}$ is limited to $\max|J_{29}|^2 \approx 4.7\%$ of the initial intensity. In practice we get around 2.9% due to back reflections (in $y$) and slightly different effective coupling along $y$ for each $k_x$ component. In the simulation we assume that the structure is periodic in $x$ with a period of 32 waveguides. The output facet is chosen to support maximal total power at the imaging row. Figure 7(c) uses the same system as Fig

7(b), changing only the size of $\eta_0$ such that Eq. (8) is no longer fulfilled. We simulate the propagation of the same wavepacket in this (non-imaging) structure up to a distance that gives maximal cross-correlation between the input and the output, which occurs at $z = 926$ µm, yet it is clear that the input and output wavepacket do not overlap. So essentially – with proper design of this simple gauge-based imaging system we can transfer an arbitrary wavepacket from an initial row at the input facet of the structure to a preselected row at the output facet. This idea works well as long as the Bloch-modes spectrum in $x$ never projects onto evanescent modes (generated by TIR) throughout propagation. This example, albeit simple, serves to demonstrate that it is possible to construct various optical devices and systems by engineering the artificial gauge fields, using just one dielectric material.

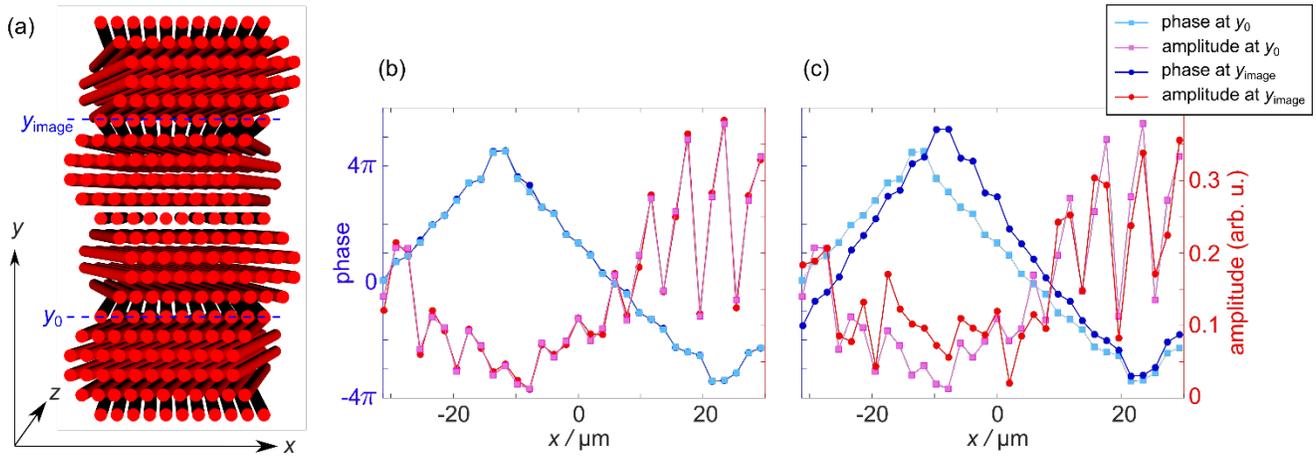

**Figure 7: Gauge-based imaging system**: (a) Sketch of a gauge-based imaging system. The tilting angle $\eta$ for each row as function of $y$ is given by $\eta(y) = \eta_0 \sin\left(\frac{2\pi(y-y_0)}{y_{\text{image}}-y_0}\right)$. We launch the beam at $y_0$ and expect it to reconstruct at $y_{\text{image}}$. In the calculations we assumed periodicity in $x$ after 32 waveguides. (b-c) amplitude (pink) and phase (light blue) at $y = y_0$ and $z = 0$ (squares) compared to amplitude (red) and phase (dark blue) at $y = y_{\text{image}}$ (circles). (b) The system parameters satisfy Eq. (8) and a reconstruction is achieved in both amplitude and phase. The output distance $z$ is chosen such that it has maximal total power at the image plane (which occurs at $z = 1000$ µm). (c) The system parameters are the same as in (b) except of $\eta_0$, such that now it does not satisfy Eq. (8). Even at the $z$ with the best fit by cross correlation ($z = 926$ µm) the original signal differs completely from the output signal.

To summarize, we derived the laws of refraction and reflection at an interface between two regions differing solely by their artificial gauge field, and demonstrated the concepts in experiments in 3D-micro printed optical waveguide arrays. Generalizing the concepts of refraction and reflection at a gauge interface offers exciting possibilities for routing light and more generally for constructing photonic systems in a given medium, strictly by designing the local gauge. As an example, we proposed an imaging system that maps any input state from one place at the input facet to a predesignated other location at the output facet, by cascading different artificial gauge fields.

**Acknowledgements**
G. v. F. acknowledges support by the Deutsche Forschungsgemeinschaft through CRC/Transregio 185 OSCAR (project number 277625399). M.S. gratefully acknowledges support by an ERC Advanced Grant, by the Israel Science Foundation, and by the German-Israel DIP project.

**Author contributions**:
All authors contributed significantly to this work.

**Additional information**
Supplementary Information accompanies this paper at http://www.nature.com/lsa/

**Competing interests**
The authors declare that they have no competing interests.

**Data and materials availability**
All experimental data and any related experimental background information not mentioned in the text are available from the authors on reasonable request.

**Correspondence**
Correspondence and requests for materials should be addressed to msegev@technion.ac.il